\documentclass[twocolumn,showpacs, reprint,aps,prl]{revtex4-1}
\usepackage{graphicx}
\usepackage{subfigure}
\usepackage{rotating}
\usepackage{amsmath, amssymb, mathtools}
\usepackage{xspace}
\usepackage{titlesec}
	\titleformat{\section}[runin]{\normalfont\normalsize\bfseries}{\thesection}{0pt}{}[.]
	\titlespacing{\section}{0pt}{\parskip+0.5ex}{2ex}

\newcommand{\ti}{t_{\mathrm{\iota}}}

\newcommand{\tf}{t_{\mathrm{f}}}

\begin{document}

%\preprint{}

\title{On the efficiency of heat engines at the micro-scale and below}
\date{\today}

\author{Paolo Muratore-Ginanneschi}
\email{paolo.muratore-ginanneschi@helsinki.fi}
\author{Kay Schwieger}
\email{kay.schwieger@helsinki.fi}
\affiliation{University of Helsinki, Department of Mathematics and Statistics
    P.O. Box 68 FIN-00014, Helsinki, Finland}

\begin{abstract}  We  investigate   the  thermodynamic  efficiency  of
sub-micro-scale heat  engines operating  under the conditions  described by
over-damped stochastic  thermodynamics. We prove that  at maximum power
the efficiency  obeys for constant isotropic mobility the 
universal  law $\eta=2\,\eta_{C}/(4-\eta_{C})$ where $\eta_{C}$ is the efficiency 
of an ideal Carnot cycle. The corresponding power optimizing protocol  
is specified by the solution  of an optimal mass  transport  problem. 
Such solution can be determined explicitly using well known 
Monge--Amp\`ere--Kantorovich reconstruction algorithms.  Furthermore,  
we  show  that the  same  law describes the efficiency of heat engines 
operating at maximum work over short time periods. Finally, we illustrate
the straightforward extension of these results to cases when the mobility 
is anisotropic and temperature dependent.
\end{abstract}

%05.40.-a: Fluctuation phenomena, random processes, noise, and Brownian motion
%05.70.Ln: Nonequilibrium and irreversible thermodynamics
%02.30.Yy Control theory
%02.50.Ey: Stochastic processes  

\maketitle

%\section{Introduction}

Present-day  technology  makes  possible  high  accuracy  handling  of
colloidal  micro  and  sub-micro  particles  to  study  experimentally
thermodynamic  processes  out  of  equilibrium  \cite{CiGoSoPe13}.   A
recent experiment, \cite{BlBe11} used a highly focused infra-red laser
beam  to  confine  a  $3\mu  m$ diameter  colloidal  particle  into  a
parabolic potential  of tunable stiffness. By  alternately varying the
stiffness  of the  potential  and,  on much  faster  time scales,  the
temperature  of   the  solvent   where  the  particle   was  suspended
\cite{BlBe11} provides  a clear evidence of  the possibility
to construct a Stirling heat engine at the micro-scale. This experiment
is ground-breaking because it opens the way of exploring the limits to
which it is possible to scale down heat engines while retaining the same
working principles of their macroscopic counter-parts. 

In order to set the scene for the present discussion it is expedient to
start by recalling the protocol governing an ideal Stirling cycle. 
During such a cycle, a working system undergoes a sequence of 
four  thermodynamic processes while being alternately in contact with two equilibrium 
reservoirs at different temperatures. First, an isothermal heat uptake $\mathcal{Q}_{in}$ 
from the environment at temperature (measured in energy units) $\beta_{h}^{-1}$.   During this process the
system expands, i.e., it converts heat  into work done on the environment.
Second,   an  isochoric   (constant-volume) cooling of the environment
to a temperature  $\beta_{c}^{-1}\,<\,\beta_{h}^{-1}$.    Third, an isothermal compression during
which the working system transfers an amount $\mathcal{Q}_{out}$ of heat to 
the cold reservoir.
Finally,    an   isochoric    heating    of    the   environment back to
$\beta_{h}^{-1}$. The neat effect of the cycle is the extraction of an
amount of work $\mathcal{W}_{out}$ from the system with efficiency measured
by  the ratio  $\eta=\mathcal{W}_{out}/\mathcal{Q}_{in}$. General  thermodynamic
reasoning  \cite{Fer56} proves  that  the efficiency  of any  heat
engine cyclically  working between two  classical equilibrium heat baths 
cannot  exceed Sadi Carnot's bound
\begin{eqnarray}
\label{Ce}
\eta_{C}=1-\frac{\beta_{h}}{\beta_{c}}
   \end{eqnarray}  
The bound is attained by performing a reversible cycle. Thus, Carnot's
bound can be classically attained in the same adiabatic limit when the power 
output tends to zero. 

Coming back to the experimental conditions implemented in \cite{BlBe11},
a distinctive property of micro and sub-micro scale 
heat engines is that they operate with characteristic energies of the order 
of thermal fluctuations  \cite{BuLiRi05,KaLeZe07}.  The state $\boldsymbol{\zeta}_{t}$
of the working system at time  $t$ must be thus characterized by a probability
distribution    $\mathtt{m}(\boldsymbol{z},t)\mathrm{d}\boldsymbol{z}=
\operatorname{P}(\boldsymbol{z}<\boldsymbol{\zeta}_{t}\leq
\boldsymbol{z}+\mathrm{d}\boldsymbol{z})$ over  the space  $\Omega$ of
admissible  configurations.  The expansion (compression) of the system 
corresponds then to  the increase (decrease)  of the fluctuations of  the particle
induced by  varying the temperature of  the solvent. Thermodynamic
indicators such as heat and work also become fluctuating quantities. 
The analysis of thermodynamic cycles needs to be rephrased in the language of
stochastic thermodynamics (see e.g. \cite{Sekimoto}, the more recents \cite{PoVeEs15} 
and refs. therein).

The micro-scale heat engine of \cite{BlBe11} relies on the heat cycle 
theoretical model \cite{ScSe08} to extract thermodynamic quantities from 
the measurements of the colloidal particle position. In \cite{ScSe08} 
the dynamics of the micro-scale particle is modeled by Langevin--Smoluchowski 
(over-damped)  equations.  Quantitative analysis  of the Stirling  cycle is
then accomplished in  the special  but  important  case   of  Gaussian
fluctuations.  The cycle starts  at time  $\ti$ when  the system
statistics is a normal distribution centered at the  origin and with
\emph{assigned}  variance  $\sigma_{a}^{2}$. Isothermal  expansion  at
temperature $\beta_{h}^{-1}$ is enacted by requiring that at a further
time $t_{o}^{-}$ the system  probability be still normally distributed
with    zero    average     but    with    \emph{assigned}    variance
$\sigma_{b}^{2}\,>\,\sigma_{a}^{2}$.  Then   the  probability  remains
continuous across  the discontinuity of the  temperature profile which
at $t=t_{o}$ drops to  $\beta_{c}^{-1}$. Isothermal compression brings
back  the   variance  of  the  distribution   to  $\sigma_{a}^{2}$  at
$t=\tf^{-}$.  Finally, the  cycle  closes at  $t=\tf$  with an  abrupt
temperature transition  from $\beta_{c}^{-1}$ to  $\beta_{h}^{-1}$. In
order to yield a finite power output  the cycle operates in a
finite   period   $\mathcal{T}=\tf-\ti$.    Hence   a   natural
definition of optimal efficiency \cite{CuAh75}  is 
that of the protocol maximizing the  power output over the duration of
the isothermal expansions and the period $\mathcal{T}$. The result for
the Gaussian cycle is \cite{ScSe08}
\begin{eqnarray}
\label{emp}
\eta_{\star}=\frac{2\,\eta_{C}}{4-\eta_{C}}
\end{eqnarray}
More qualitative considerations led \cite{ScSe08} to suggest for non-Gaussian
statistics and anisotropic temperature-dependent mobility
\begin{eqnarray}
\label{quali}
\eta_{\star}=\frac{\eta_{C}}{2-\alpha\,\eta_{C}}
\end{eqnarray} 
with $0\leq \alpha\leq 1$ admitting an explicit expression only in the case of isotropic mobility and 
recovering (\ref{emp}) for constant mobility. 
In order to compare these predictions with experiments, two
questions were left open. First, to derive in the most general case the value of $\alpha$ 
in terms of the parameters of the over-damped dynamics and, second and most importantly, to
prove the \emph{realizability} of (\ref{emp}) and (\ref{quali}). This means to determine
the protocols for which (\ref{emp}), (\ref{quali}) are attained.
A theoretical effort in this direction was explicitly demanded by the authors 
of the micro-scale heat engine \cite{BuLiRi05}.

In the  present contribution we address both questions. We specify in formula (\ref{alpha}) 
below the value of the constant $\alpha$ appearing in (\ref{quali}) in terms of the
mobility for any reasonable probability distributions describing 
the state of the system at the end of the isothermal processes. Furthermore, 
we prove that power optimizing protocol is specified by the    well-known
Monge--Amp\`ere--Kantorovich     mass     reconstruction     algorithm
\cite{BeBr00,BrFrHeLoMaMoSo03}. To neaten the notation, we report in details
the calculations in the case of constant isotropic mobility, the extension to
the general case being straightforward.

We  prove also  a  second  universality result for the efficiency.  
It  was noticed  in  \cite{Lef87}  that the efficiency at 
maximum power for a Stirling cycle well approximates also the efficiency at 
maximum work output in the limit of short duration of the cycle even for 
ideal heat cycles other than Stirling's.  Indeed, we show  that if we optimize
the   work   output   with   respect  to   the   ``expansion''   state
$\mathtt{m}_{o}$  and to the duration of the isothermal expansion then, 
in  the limit  of short  cycle period ($\mathcal{T}\downarrow 0$), the efficiency 
tends to (\ref{quali}) . We  expect this  result to  be relevant  for the
design of  nano-scale engines  when controlling  ``target expansion''
state $\mathtt{m}_{o}$ may not be feasible.

Let us briefly  explain the origin of the universality of our results.  
It  stems from the fact that the entropy production during an arbitrary 
thermodynamic transition is proportional to the kinetic energy  associated 
to the current velocity \cite{Nelson01} of the system. This fact  was known 
for some time (see e.g.  \cite{JiangQianQian}). Its  central  role,  however, 
in  mapping optimal control  of stochastic thermodynamic transitions  into optimal
deterministic   mass  transport   \cite{Vil03,DePhFi14}  was   only  recently
understood  \cite{AuMeMG12} (see  also \cite{PMG13})  and subsequently
applied to predict refinements to  the second law of thermodynamics and
to Landauer's bound \cite{AuGaMeMoMG12} which have found experimental
confirmations see e.g. \cite{CiGoSoPe13}.

\section{Model}

We suppose that the configuration space $\Omega$ of working system is $\mathbb{R}^{d}$,
while its state evolves according to the Langevin--Smoluchowski dynamics
\begin{eqnarray}
\label{model:eqs1}
\mathrm{d}\boldsymbol{\zeta}_{t}=-\partial_{\boldsymbol{\zeta}_{t}}U(\boldsymbol{\zeta}_{t},t)\,
\mathrm{d}t+\sqrt{ \frac{2}{\beta_{t}}}\mathrm{d}\boldsymbol{\omega}_{t}
\end{eqnarray}
The system is driven by the gradient of a function $U\colon\mathbb{R}^{d}\times\mathbb{R}\mapsto\mathbb{R}$ 
which we take in the class of confining potentials sufficiently regular to justify the 
manipulations which follow. Let us also start by taking for $\beta_{t}^{-1}$ 
a regularized, $\mathcal{T}$-periodic and differentiable, version of the temperature time profile used 
in \cite{ScSe08}.  According to the general framework of stochastic thermodynamics 
\cite{Se98,Qia01}, the mean heat dissipated by the working system during a 
temperature cycle is equal to the expectation value of the \emph{Stratonovich stochastic integral}
\begin{eqnarray}
\label{heat}
\mathcal{Q}=-\operatorname{E}\int_{\ti}^{\tf}\mathrm{d}\boldsymbol{\zeta}_{t}
\cdot\partial_{\boldsymbol{\zeta}_{t}}U(\boldsymbol{\zeta}_{t},t)
\end{eqnarray}
Well known properties of the Stratonovich integral yield after elementary manipulations 
\cite{AuMeMG12}
\begin{eqnarray}
\label{heat2}
&&\mathcal{Q}=\beta_{\ti}^{-1}\,\mathcal{S}_{\ti}-\beta_{\tf}^{-1}\,\mathcal{S}_{\tf}+
\nonumber\\
&&\operatorname{E}\int_{\ti}^{\tf}\mathrm{d}t \left\{\|\boldsymbol{v}(\boldsymbol{\zeta}_{t},t)\|^{2}
+S(\boldsymbol{\zeta}_{t},t)\frac{\mathrm{d} }{\mathrm{d} t}\frac{1}{\beta_{t}}\right\}
\end{eqnarray}
where $\boldsymbol{v}(\boldsymbol{z},t)=
-\partial_{\boldsymbol{z}}\left(U(\boldsymbol{z},t)+\beta_{t}^{-1}\,S(\boldsymbol{z},t)\right)$
is the current velocity of the system, $S(\boldsymbol{z},t)=-\ln [\mathtt{m}(\boldsymbol{z},t)/C]$ 
the microscopic entropy and $\mathcal{S}_{t}=\operatorname{E}S(\boldsymbol{\zeta}_{t},t)$.  $C$ is a dimensional constant 
irrelevant for the considerations which follow.
From (\ref{heat2}) we immediately see that 
if $\beta_{\ti}=\beta_{\tf}$, $\mathtt{m}(\boldsymbol{z},\ti)=\mathtt{m}(\boldsymbol{z},\tf)$
the only contribution to the dissipated heat comes from the entropy production. 
Owing to the non-degeneracy of the noise in the over-damped approximation, we are free 
to control the entropy production in terms of the current velocity $\boldsymbol{v}$.
Optimal control strategies are then most conveniently found by searching for extremals 
with respect to $\mathtt{m}$ and $\boldsymbol{v}$ of the Pontryagin--Bismut functional \cite{KoPa93} 
\begin{eqnarray}
\label{}
&&\mathcal{A}=\operatorname{E}[V(\boldsymbol{\zeta}_{tf},\tf)-V(\boldsymbol{\zeta}_{\ti},\ti)]+
\nonumber\\&&
\operatorname{E}\int_{\ti}^{\tf}\hspace{-0.2cm}\mathrm{d}t \left\{\|\boldsymbol{v}(\boldsymbol{\zeta}_{t},t)\|^{2}
+S(\boldsymbol{\zeta}_{t},t)\frac{\mathrm{d} }{\mathrm{d} t}\frac{1}{\beta_{t}}
-(\mathfrak{D}V)(\boldsymbol{\zeta}_{t},t)\right\}
\end{eqnarray}
The co-state function $V\colon\mathbb{R}^{d}\times\mathbb{R}\mapsto\mathbb{R}$ plays here the role 
of a Lagrange multiplier imposing that
\begin{eqnarray}
\label{}
(\mathfrak{D}V)(\boldsymbol{z},t)=
[\partial_{t}+\boldsymbol{v}(\boldsymbol{z},t)\cdot\partial_{\boldsymbol{z}}]V(\boldsymbol{z},t)
\end{eqnarray}
acts on scalars as a total derivative along the flow generated by $\boldsymbol{v}$.

\section{Efficiency at maximum power} 

We construct the optimal cycle by optimizing (\ref{heat2}) \emph{separately} in
$[\ti\,,t_{o})$ and $[t_{o}\,,\tf)$ under the boundary conditions 
$\mathtt{m}(\boldsymbol{z},\ti)=\mathtt{m}(\boldsymbol{z},\tf)=\mathtt{m}_{\iota}(\boldsymbol{z})=C\,
\exp\left\{-S_{\iota}(\boldsymbol{z})\right\}$
and $\mathtt{m}(\boldsymbol{z},t_{o})=\mathtt{m}_{o}(\boldsymbol{z})=C\,
\exp\left\{-S_{o}(\boldsymbol{z})\right\}$. We choose $S_{o}$ and $S_{\iota}$ such that the probability
densities are smooth, and that 
the Gibbs-Shannon entropy
\begin{eqnarray}
\label{}
\mathcal{S}_{t_{o}}-\mathcal{S}_{\ti}\equiv\mathcal{S}_{o}-\mathcal{S}_{\iota}\equiv
\operatorname{E}[S_{o}(\boldsymbol{\zeta}_{t_{o}})-S_{\iota}(\boldsymbol{\zeta}_{\ti})]\geq 0
\end{eqnarray}
to signify an expansion of the system. In the limit of abrupt
temperature changes, the derivative of the temperature vanishes except for Dirac-$\delta$
contributions localized at $t_{o}$ and $\tf$. Proceeding as in \cite{AuMeMG12,AuGaMeMoMG12}, 
we find that along the isothermal branches of the
cycle the optimal current velocity satisfies the two equations
\begin{eqnarray}
\label{oceqs}
\mathfrak{D}\boldsymbol{v}=0\hspace{0.5cm}\&\hspace{0.5cm}\mathfrak{D}S-\partial_{\boldsymbol{z}}\cdot\boldsymbol{v}=0
\end{eqnarray}
complemented by the stationarity condition $\partial_{\boldsymbol{z}}V=2\,\boldsymbol{v}$. The set of these three 
equations define the Monge--Amp\`ere--Kantorovich optimal mass transport problem
\cite{Vil03,DePhFi14}.  Let us focus on the sub-interval $[\ti,t_{o})$.
For any $t\in[\ti,t_{o})$  the velocity satisfies 
\begin{eqnarray}
\label{ocv}
\boldsymbol{v}(\boldsymbol{\phi}(t;\boldsymbol{z},\ti),\ti)=\boldsymbol{v}(\boldsymbol{z},\ti)
\equiv\frac{1}{2}\partial_{\boldsymbol{z}}V(\boldsymbol{z},\ti)
\end{eqnarray}
with free steaming characteristics
\begin{eqnarray}
\label{lm}
\boldsymbol{\phi}(t;\boldsymbol{z},\ti)=\boldsymbol{z}+\,\frac{1}{2}\partial_{\boldsymbol{z}}V(\boldsymbol{z},\ti)(t-\ti)
\end{eqnarray}
The initial velocity is on its turn determined by the solution of the 
boundary problem 
\begin{eqnarray}
\label{bp}
\lefteqn{
S(\boldsymbol{z},t_{o})
}
\nonumber\\&&
=
S(\boldsymbol{\phi}(t_{o};\boldsymbol{z},\ti),t_{o})-
\operatorname{tr}\ln [\partial_{\boldsymbol{z}}\otimes\boldsymbol{\phi}(t_{o};\boldsymbol{z},\ti),t_{o})]
\end{eqnarray}  
In writing (\ref{bp}) we imposed the continuity in time of the probability measure 
and of the Lagrangean map (\ref{lm}) at $t_{o}$. The crucial observation is that since
$S(\boldsymbol{z},t_{o})=S_{\iota}(\boldsymbol{z})$,
$S(\boldsymbol{\phi}(t_{o};\boldsymbol{z},\ti),t_{o})=S_{o}(\boldsymbol{\phi}(t_{o};\boldsymbol{z},\ti))$ 
are assigned, the \emph{Lagrangian map} 
$\boldsymbol{\phi}(t_{o};\boldsymbol{z},\ti)=\boldsymbol{\phi}_{\star}(\boldsymbol{z})$ 
depends only on the boundary conditions and \emph{not} on the duration of the isothermal process.
As we can repeat the same considerations for the optimization in $[t_{o},\tf)$, 
we arrive after elementary manipulations (see e.g. \cite{AuGaMeMoMG12} for details) at the 
general expression of the heat dissipated over one cycle
\begin{eqnarray}
\label{}
\mathcal{Q}=\left(\frac{1}{\beta_{h}}-\frac{1}{\beta_{h}}\right)(\mathcal{S}_{\iota}-\mathcal{S}_{o})+
\frac{\gamma\,(1-\gamma)}{\mathcal{T}}\mathcal{K}
\end{eqnarray}
For convenience we introduced the ratio $\gamma=(t_{o}-\ti)/\mathcal{T}\in [0,1]$ and defined
\begin{eqnarray}
\label{}
\mathcal{K}=\operatorname{E}\|\boldsymbol{\phi}_{\star}(\boldsymbol{\zeta}_{\ti})-\boldsymbol{\zeta}_{\ti}\|^{2}\geq 0
\end{eqnarray}
From the mathematical slant, $\mathcal{K}$ is the squared Wasserstein distance
between the probability measures specified by $\mathtt{m}_{\iota}(\boldsymbol{z})$
and $\mathtt{m}_{o}(\boldsymbol{z})$ \cite{Gaw13}. 
Following \cite{ScSe08} we define the heat input during the cycle as
\begin{eqnarray}
\label{hi}
\mathcal{Q}_{in}=\frac{\mathcal{S}_{o}-\mathcal{S}_{\iota}}{\beta_{h}}-\frac{\mathcal{K}}{\gamma\,\mathcal{T}}
\end{eqnarray}
and the heat output
\begin{eqnarray}
\label{ho}
\mathcal{Q}_{out}=-\frac{\mathcal{S}_{o}-\mathcal{S}_{\iota}}{\beta_{c}}
-\frac{\mathcal{K}}{(1-\gamma)\,\mathcal{T}}
\end{eqnarray}
The definitions imply $-\mathcal{Q}=\mathcal{Q}_{in}+\mathcal{Q}_{out}=\mathcal{W}_{out}$
where $\mathcal{W}_{out}$ is the work output.
The power of the cycle is then
\begin{eqnarray}
\label{}
\wp\equiv\frac{\mathcal{W}_{out}}{\mathcal{T}}
=\eta_{C}\frac{\mathcal{S}_{o}-\mathcal{S}_{\iota}}{\beta_{h}\,\mathcal{T}}
-\frac{\mathcal{K}}{\gamma\,(1-\gamma)\,\mathcal{T}^{2}}
\end{eqnarray}
and its efficiency
\begin{eqnarray}
\label{oef}
\eta=\frac{\mathcal{W}_{out}}{\mathcal{Q}_{in}}=1
-\frac{(1-\eta_{C})\,(\mathcal{S}_{o}-\mathcal{S}_{\iota})
+\frac{\beta_{h}\,\mathcal{K}}{(1-\gamma)\,\mathcal{T}}}
{\mathcal{S}_{o}-\mathcal{S}_{\iota}-\frac{\beta_{h}\,\mathcal{K}}{\gamma\,\mathcal{T}}}
\end{eqnarray}
having used $1-\eta_{C}\equiv\beta_{h}\,\beta_{c}^{-1}$.
The maximum power is then attained for $\gamma_{\star}=1/2$ and
\begin{eqnarray}
\label{}
\mathcal{T}_{\star}=\frac{8\,\beta_{h}\,\mathcal{K}}{\eta_{C}\,(\mathcal{S}_{o}-\mathcal{S}_{\iota})}
\end{eqnarray}
and it is equal to
\begin{eqnarray}
\label{}
\wp_{\star}=\frac{\eta_{C}^{2}\,(\mathcal{S}_{o}-\mathcal{S}_{\iota})^{2}}{16\,\mathcal{K}\,\beta_{h}^{2}}
\end{eqnarray}
Finally evaluating the efficiency at maximum power $\gamma=\gamma_{\star}$,
$\mathcal{T}=\mathcal{T}_{\star}$ yields (\ref{emp}) independently
of $\mathtt{m}_{\iota}$ and $\mathtt{m}_{o}$. The protocol attaining the maximum power is specified
by (\ref{bp}), which can be numerically solved using the algorithms
given in \cite{BeBr00,BrFrHeLoMaMoSo03}. This is the first
of our announced universality results.

\section{Efficiency at maximum work output}

We now consider a different optimization setting. As before, we allow the temperature to
change only at $t_{o}$ and $\tf$. Also as before we assign 
$\mathtt{m}(\boldsymbol{z},\ti)=\mathtt{m}(\boldsymbol{z},\tf)=\mathtt{m}_{\iota}(\boldsymbol{z})=C\,
\exp\left\{-S_{\iota}(\boldsymbol{z})\right\}$ and we wish to optimize with respect to the duration
of the isothermal processes. The difference is now that we look for minima
of the dissipated heat (equivalently, maxima of the work output) with respect to
$\mathtt{m}(\boldsymbol{z},t_{o})$.
 
In such a case, the optimal current velocity still obeys (\ref{oceqs}) except when 
the isochoric temperature changes occur. Upon requiring the continuity of the
probability density at $t_{o}$ we find
\begin{eqnarray}
\label{}
\boldsymbol{v}(\boldsymbol{z},t_{o})-\boldsymbol{v}(\boldsymbol{z},t_{o}^{-})=
\eta_{C}\,\frac{(\partial_{\boldsymbol{z}}S)(\boldsymbol{z},t_{o})}{2\,\beta_{h}}
\end{eqnarray}
We can then avail us of the continuity of the current velocity characteristics 
at $t_{o}$, their free streaming form along isothermal processes, and of (\ref{ocv}) 
to derive the extremal condition for the value of the initial velocity
\begin{eqnarray}
\label{}
\boldsymbol{v}(\boldsymbol{z},\ti)
=(1-\gamma)\,\eta_{C}
\,\frac{(\partial_{\boldsymbol{\phi}}S)(\boldsymbol{\phi}(t_{o};\boldsymbol{z},\ti),t_{o})}{2\,\beta_{h}}
\end{eqnarray}
Combining this equation with (\ref{bp}) we finally get into a 
closed equation for the Lagrangian map $\varphi(\boldsymbol{z})\equiv\boldsymbol{\phi}(t_{o};\boldsymbol{z},\ti)$
governing the maximum work output
\begin{eqnarray}
\label{mwoeq}
\hspace{-0.5cm}
\frac{2\,\beta_{h}\,(\partial_{\boldsymbol{z}}\otimes\boldsymbol{\varphi})}{\gamma\,(1-\gamma)\,\eta_{C}}
\cdot\frac{\boldsymbol{\varphi}-\boldsymbol{z}}{\mathcal{T}}
-\partial_{\boldsymbol{z}}\operatorname{tr}\ln(
\partial_{\boldsymbol{z}}\otimes\boldsymbol{\varphi})
=\partial_{\boldsymbol{z}}S_{\iota}
\end{eqnarray}
We complement this equation with the boundary condition that 
$\varphi$ maps the support of the initial density into itself
($
\boldsymbol{\varphi}(\mathbb{R}^{d})=\mathbb{R}^{d}
$). We emphasize that the stationarity condition 
$2\,\boldsymbol{v}=\partial_{\boldsymbol{z}}V$ implies that the Lagrangian map $\boldsymbol{\varphi}$
is itself a gradient map. Hence, under suitable regularity hypotheses, we have reason 
to expect that the problem (\ref{mwoeq}) is  well posed \cite{DePhFi14}.
We also observe that $\partial_{\gamma}\boldsymbol{\varphi}=0$ generically solves the variational 
equation obtained by differentiating (\ref{mwoeq}) with respect to $\gamma$ if $\gamma=1/2$.
In other words, solutions of (\ref{mwoeq}) attain stationarity with respect to $t_{o}$
if the isothermal expansion takes half of the cycle period. This result is intuitive in light of the
fact that we could optimize the work output by first looking for the optimal duration of the expansion 
for fixed target state at $t_{o}$ and then search for the optimal state. Finally, differentiating
(\ref{mwoeq}) with respect to $\mathcal{T}$ indicates that we cannot generically expect optimization with
respect to $\mathcal{T}$ to be possible. We therefore conclude that the solution $\boldsymbol{\varphi}_{\star}$ 
of (\ref{mwoeq}) describing maximum work output is obtained at $\gamma=1/2$ and depends parametrically 
upon $\mathcal{T}$.

We now are interested in solving (\ref{mwoeq}) in the limit of very short cycle period. 
As (\ref{lm}) holds up to $t_{o}$, we obtain for $\mathcal{T}$ and $\boldsymbol{z}$
sufficiently small
\begin{eqnarray}
\label{mwoapprox}
\varphi_{\star}\approx\boldsymbol{z}+\frac{\eta_{C}\,\mathcal{T}}{8\,\beta_{h}}\partial_{\boldsymbol{z}}S_{\iota}
\end{eqnarray}
If the initial value of the microscopic entropy $S_{\iota}$ confines fluctuations in the bulk
region where (\ref{mwoapprox}) holds, we are then in the position to compute the efficiency of the
cycle in short period limit. Namely, straightforward manipulations show that also in the present 
case the heat input and output are respectively amenable to the form (\ref{hi}) and (\ref{ho}).
Using (\ref{mwoapprox}) we obtain within accuracy the expression of the Shannon-Gibbs entropy
\begin{eqnarray}
\label{}
\lefteqn{
\operatorname{E}\operatorname{tr}\ln\left(\partial_{\boldsymbol{\zeta}_{\ti}}
\otimes\boldsymbol{\varphi}_{\star}\right)(\boldsymbol{\zeta}_{\ti})
}
\nonumber\\&&
\approx \frac{\eta_{C}\,\mathcal{T}}{8\,\beta_{h}}
\int_{\mathbb{R}^{d}}\mathrm{d}^{d}\boldsymbol{z}\,C\,e^{-S_{\iota}}\partial_{\boldsymbol{z}}^{2}S_{\iota}
=\frac{\eta_{C}\,\mathcal{T}}{8\,\beta_{h}}\operatorname{E}\|\partial_{\boldsymbol{\zeta}_{\ti}}S_{\iota}\|^{2}
\end{eqnarray}
Similarly, we obtain for the squared Wasserstein distance the estimate
\begin{eqnarray}
\label{}
\mathcal{K}\approx\left(\frac{\eta_{C}\,\mathcal{T}}{8\,\beta_{h}}\right)^{2}
\operatorname{E}\|\partial_{\boldsymbol{\zeta}_{\ti}}S_{\iota}\|^{2}
\end{eqnarray}
Upon inserting these results into the expression of the efficiency we obtain
\begin{eqnarray}
\label{eff:Leff}
\lim_{\mathcal{T}\downarrow 0}\eta_{\star}=\frac{2\,\eta_{C}}{4-\eta_{C}}
\end{eqnarray}
We have therefore validated in the framework of overdamped stochastic thermodynamics the
result of \cite{Lef87}. Accordingly, the efficiency of work at maximum output 
can be approximated for small cycle period and small temperature ratio by same 
formula describing efficiency at maximum power.

\section{Elementary Extensions}

In principle, anisotropic effects in the solvent may require to consider 
\begin{eqnarray}
\label{model:aniso}
\mathrm{d}\boldsymbol{\zeta}_{t}=-\mathsf{M}_{t}\cdot\partial_{\boldsymbol{\zeta}_{t}}U(\boldsymbol{\zeta}_{t},t)\,
\mathrm{d}t+\sqrt{ \frac{2\,\mathsf{M}_{t}}{\beta_{t}}}\cdot\mathrm{d}\boldsymbol{\omega}_{t}
\end{eqnarray}
Here $\mathsf{M}_{t}$ is a strictly positive definite symmetric tensor
modeling the mobility of the solvent. For many experimental applications, 
it is adequate to assume that the mobility tensor is constant in space and
depends on time only because of changes in the solvent temperature: $\mathsf{M}_{t}\equiv\mathsf{M}(\beta_{t})$.
Under these hypotheses, we can repeat step-by-step the above calculations. 
The relevant indicators along isothermal transformations become the square 
Wasserstein distances :
\begin{eqnarray}
\label{}
\mathcal{K}_{\ell}=\operatorname{E}\left\{(\boldsymbol{\phi}_{\star}(\boldsymbol{\zeta}_{\ti})-\boldsymbol{\zeta}_{\ti})
\cdot\mathsf{M}_{\ell}^{-1}\cdot(\boldsymbol{\phi}_{\star}(\boldsymbol{\zeta}_{\ti})-\boldsymbol{\zeta}_{\ti})\right\}
\end{eqnarray}
where $\mathsf{M}_{\ell}\equiv\mathsf{M}(\beta_{\ell})$, $\ell=h,c$ and $\boldsymbol{\phi}_{\star}$
is either the solution of the Monge--Amp\`ere--Kantorovich equation
or of the counter-part of (\ref{mwoeq}):
\begin{eqnarray}
\label{}
\hspace{-0.5cm}
\frac{2\,\beta_{h}\,(\partial_{\boldsymbol{z}}\otimes\boldsymbol{\varphi})}{\eta_{C}\,\gamma\,(1-\gamma)}
\cdot
\mathsf{C}
\cdot
\frac{\boldsymbol{\varphi}-\boldsymbol{z}}{\mathcal{T}}
-
\partial_{\boldsymbol{z}}\operatorname{tr}\ln(\partial_{\boldsymbol{z}}\otimes\boldsymbol{\varphi})
=\partial_{\boldsymbol{z}}S
\end{eqnarray}
for $\mathsf{C}=\gamma\,\mathsf{M}_{c}^{-1}+(1-\gamma)\,\mathsf{M}_{h}^{-1}$.
In both cases, we obtain a generalization of (\ref{emp}) and (\ref{eff:Leff}) 
amenable to the form (\ref{quali}) with
\begin{eqnarray}
\label{alpha}
0\leq\,\alpha=\frac{\sqrt{ \mathcal{K}_{h}}}{\sqrt{ \mathcal{K}_{h}}+\sqrt{ \mathcal{K}_{c}}}\leq 1
\end{eqnarray}
Using the lower and upper bounds for $\alpha$, we readiliy recover the inequalities
\begin{eqnarray}
\label{EsKaLiVdBr10}
\frac{\eta_{C}}{2}\leq \eta_{\star}\leq \frac{\eta_{C}}{2-\eta_{C}}
\end{eqnarray} 
derived in \cite{EsKaLiVdBr10} by assuming a time asymptotic behavior 
of the entropy production along isothermal processes inversely proportional
to their duration. This is not surprising because the entropy production 
by optimal protocols in over-damped thermodynamics exactly satisfies such 
condition \cite{AuGaMeMoMG12,Gaw13}.

\section{Conclusions and Perspectives}

To summarize, we derived the explicit expression (\ref{quali}) and (\ref{alpha}) 
of the efficiency at maximum power for the Stirling heat cycle in the 
framework of over-damped (Langevin--Smoluchowski) dynamics. We also 
determined the algorithm for computing the protocols achieving such
efficiency. Present day technological  advances occur at an impressive
speed.  Laboratory  evidences  of   the  possibility  of  implementing
Stirling cycles at the \emph{nano}-scale  is already documented in the
literature  \cite{KiBlDeGrKaAs13}   and  the  design  of   a  possible
experiment has been recently proposed \cite{DeKiLu15}, see also 
\cite{MaRoDiPeRi15,MaRoDiPePaRi15}.  These advances pose the
challenge  to   repeat  the   present  analysis  in   the  under-damped
(Langevin--Kramers) regime.  We recently proposed a  general theory of
optimal  control of  the dissipated  heat for  nano-mechanical systems
governed  by a  Langevin--Kramers  dynamics  \cite{PMGSc14}. There  we
showed  that  the overdamped  approximation  is  not only  useful  for
perturbative  analysis of  the Langevin--Kramers  but also  provides a
priori lower bounds on the dissipated heat. Based on these results, we
expect  that the  present contribution  will  be relevant  also for  the
analysis of nano-scale cyclic heat engines.

\section{Acknowledgments}

We acknowledge the Centre of
Excellence in Analysis and Dynamics Research (Academy
of Finland decision no. 271983) for support.

\addcontentsline{toc}{section}{Bibliography}
\bibliography{/home/paolo/RESEARCH/BIBTEX/jabref}{} 
\bibliographystyle{aipauth4-1} %base bibstyle abbrv.bst with eprint field

\end{document}